\documentclass[12pt]{article}
\usepackage{amsmath,amssymb,graphicx} 
\usepackage{epsf}
\usepackage{epsfig}
\usepackage{pstricks}
\usepackage{axodraw}
\usepackage{cite}

\newcommand{\met}{\not\!\!\!E_T}

\newcommand{\be}{\begin{equation}}
\newcommand{\ee}{\end{equation}}

\newcommand{\beq}{\begin{eqnarray}}
\newcommand{\eeq}{\end{eqnarray}}

\newcommand{\centeron}[2]{{\setbox0=\hbox{#1}\setbox1=\hbox{#2}\ifdim
\wd1>\wd0\kern.5\wd1\kern-.5\wd0\fi
\copy0

\kern-.5\wd0\kern-.5\wd1\copy1\ifdim\wd0>\wd1
                                       \kern.5\wd0\kern-.5\wd1\fi}}
\newcommand{\ltap}{\>\centeron{\raise.35ex\hbox{$<$}}
                               {\lower.65ex\hbox{$\sim$}}\>}
\newcommand{\gtap}{\>\centeron{\raise.35ex\hbox{$>$}}
                               {\lower.65ex\hbox{$\sim$}}\>}

\newcommand\ZZ{\hbox{\zfont Z\kern-.4emZ}}
\font\zfont = cmss10 

\begin{document}
\begin{titlepage}
\begin{flushright}
\end{flushright}

\vskip.5cm
\begin{center}
{\huge \bf 
The Invisible $Z'$ at the LHC
}

\vskip.1cm
\end{center}
\vskip0.2cm

\begin{center}
{\bf
Frank J. Petriello, Seth Quackenbush,
Kathryn M. Zurek}
\end{center}
\vskip 8pt

\begin{center}
{\it Physics Department, University of Wisconsin, Madison, WI 53706} \\

\vspace*{0.1cm}
\vspace*{0.3cm}
{\tt frankjp@physics.wisc.edu, squackenbush@wisc.edu, kzurek@wisc.edu}
\end{center}

\vglue 0.3truecm

\begin{abstract}
\vskip 3pt
\noindent
We study the feasibility of observing an invisibly decaying $Z'$ at the LHC through the process $pp \rightarrow Z Z' \rightarrow \ell^+\ell^- X X^\dagger$, where $X$ is any neutral, (quasi-) stable particle, whether a Standard Model (SM) neutrino or a new state.  The measurement of the invisible width through this process facilitates both a 
model independent measurement of $\Gamma_{Z' \rightarrow \bar{\nu} \nu}$ and potentially detection of light neutral hidden states.  Such particles appear in many models 
where the $Z'$ is a messenger to a hidden sector, and also if dark matter is charged under the $U(1)'$ of the $Z'$.  We find that with as few as $30 \mbox{ fb}^{-1}$ of data the invisibly decaying $Z'$ can be observed at $5 \sigma$ over SM background for a 1 TeV $Z'$ with reasonable couplings.  If the $Z'$ does not couple to leptons and therefore cannot be observed 
in the Drell-Yan channel, this process becomes a discovery mode.  For reasonable hidden sector couplings, masses up to 2 TeV can be probed at the LHC.  If the 
$Z'$ does couple to leptons, then the rate for this invisible decay is predicted by on-peak data and the presence of additional hidden states can be searched for.  With 
$100 \mbox{fb}^{-1}$ of data, the presence of excess decays to hidden states can be excluded at 95\% C.L. if they comprise 20-30\% of the total invisible cross section.
\end{abstract}

\end{titlepage}

\newpage


\section{Introduction}
\label{sec:intro}

New massive $U(1)$ gauge bosons appear in numerous theories of physics beyond the Standard Model (SM).  They appear in grand unified theories such as $SO(10)$~\cite{mohapatra} 
and $E(6)$~\cite{Hewett:1988xc}, in theories of extra space-time dimensions as Kaluza-Klein excitations of the SM gauge bosons~\cite{Hewett:2002hv}, and in 
Little Higgs theories of the 
electroweak sector~\cite{Schmaltz:2005ky}.  $Z'$ bosons that decay to leptons have a simple, clean experimental signature, and consequently can be searched for up to high 
masses at colliders.  Current direct search limits from Tevatron experiments restrict the $Z'$ mass to be greater than about 900 GeV when its couplings to SM fermions are identical to those of the $Z$ boson~\cite{Tev:2007sb}.  The LHC experiments are expected to extend the $Z'$ mass reach to more than 5 TeV~\cite{tdrs}.

Since the Z' signature is clean and its QCD uncertainties are small, it is likely that the couplings of a discovered $Z'$ can be studied with reasonable
accuracy to probe the high scale theory that gave rise to it.  Many studies of how to measure $Z'$ properties and couplings to SM particles have been performed~\cite{zprevs}.  
A recent study performed a next-to-leading order QCD analysis of $Z'$ properties at the LHC accounting for statistical, residual scale, and parton distribution error estimates, and 
concluded that four generation independent combinations of $Z'$ couplings could be extracted at the LHC by making full use of available on-peak differential spectra~\cite{Petriello:2008zr} (another recent study on searching for the $Z'$ is found in \cite{coriano}).  
However, a degeneracy between quark and lepton couplings can not be removed by studying $Z'$ bosons in the Drell-Yan channel; all observables in this mode are unchanged 
if the quark couplings are scaled by a factor $x$ while the lepton couplings are scaled by $1/x$.  A different production mechanism must be utilized to remove this degeneracy.  
Possibilities are $pp \to Z' \to jj,b\bar{b},$ and $t\bar{t}$; however, because of SM backgrounds, all three are expected to be extremely difficult to observe at the LHC~\cite{expstudies}.

Another possible way of removing this degeneracy is by using the $Z'$ width.  The width takes the form
\begin{equation}
\Gamma = \Gamma_{inv}+\Gamma_{oth}+\sum_q \Gamma_q+\sum_l \Gamma_l.
\label{wdecomp}
\end{equation}
$\Gamma_{inv}$ is the partial width for $Z'$ decays into invisible states such as SM neutrinos, $\Gamma_q$ and $\Gamma_l$ denote the widths for $Z'$ decays into quarks 
and leptons respectively, and $\Gamma_{oth}$ represents possible other decay modes such as $Z' \to W^+W^-,Zh$.  This relation does not suffer from the same degeneracy as noted above.  
The total width $\Gamma$ can be measured by fitting the shape of the resonance peak assuming the $Z'$ is not too narrow.  $\Gamma_{oth}$ is small for large classes of models.  If 
we make the mild theoretical assumption that $SU(2)_L$ invariance equates the $Z'$ couplings of charged leptons to those of neutrinos (satisfied in grand unified models), and note that the 
on-peak study of~\cite{Petriello:2008zr} showed that the combination $c_q \sim \Gamma_q \Gamma_l / \Gamma^2$ can be measured, Eq.~(\ref{wdecomp}) becomes a quadratic 
equation for the unknown $\Gamma_q$, $\Gamma_l$ that can be solved up to a two-fold discrete ambiguity.  The only other assumption entering this procedure is that 
$\Gamma_{inv}$ is composed entirely of $Z'$ decays to neutrinos.

Besides breaking this degeneracy between quark and lepton couplings, there is an additional strong motivation for studying the invisible width of the $Z'$.  $Z'$ bosons often appear as messengers which connect the SM to hidden sectors, such as in some models of supersymmetry breaking~\cite{Chung:2003fi} and in Hidden Valley models~\cite{Strassler:2006im}, and can 
decay to light particles in this hidden sector.  For example, Hidden Valley models contain sub-TeV mass states which are electrically neutral and quasi-stable, with decay lengths in some cases longer than tens of meters.  These exit the detector as missing energy.  A sterile neutrino which is charged under the $U(1)'$ would also result in hidden decays of the $Z'$.  Such states may also account for the observed dark matter, as in the model of~\cite{Hooper:2008im}.  A model of milli-charged dark matter from a Stueckelberg $Z'$ may also be found in Ref.~\cite{Cheung}.

In this paper we study whether invisible decays of the $Z'$ can be detected at the LHC using the channel $pp \to ZZ' \to \ell^+\ell^-\met$.  This mode has previously been used to search for 
invisible decays of the Higgs boson~\cite{invh}.  As we will be interested in the large missing $E_T$ kinematic region, $\met \sim 200$ GeV, the experimental signature is 
relatively clean.  Other possible channels such as $pp \to \gamma \met,j\met$ are sensitive to significant uncertainties such as jet energy mismeasurements and jets faking photons.  We demonstrate that invisible $Z'$ decays can be seen over the SM background with a significance of $S/\sqrt{B}=3$ with as little as $10\,{\rm fb}^{-1}$ 
for realistic models, while $S/\sqrt{B}=5$ can be obtained with $30\,{\rm fb}^{-1}$.  We show that the structure of the $pp \to ZZ' \to \ell^+\ell^-\met$ cross section 
admits a simple parametrization using two effective charges, associated with emission of the $Z$ boson from either intitial state quarks or final state neutrinos.  This 
allows invisible $Z'$ decays to be studied in a model-independent fashion.  For hidden sector states, only the initial state radiation contribution occurs.  
Assuming that the only invisible decays of the $Z'$ are to SM neutrinos, these 
charges are predicted by the Drell-Yan study in~\cite{Petriello:2008zr}.  Any deviation would indicate $Z'$ couplings to light hidden sector states.  We quantify 
what deviations can be seen given expected errors.  We find that hidden sector decays making up 20-30\% of the total invisible width can be observed at the LHC.  If 
the $Z'$ does not couple to leptons but decays to hidden sector states, $pp \to ZZ' \to \ell^+\ell^-\met$ becomes a discovery mode.  We show that leptophobic 
$Z'$ bosons with masses up to 2 TeV can be probed at the LHC.  With an integrated luminosity of $100\,{\rm fb}^{-1}$, one can exclude a pure hidden sector $Z'$ with $\sigma_{hid} > 0.3$ fb with a confidence of $95\%$; for $1000\,{\rm fb}^{-1}$, one can exclude decays to hidden sector states down to $0.1$ fb.  A $3 \sigma$ discovery can be achieved with $\sigma_{hid} > 0.6$ fb for $100\,{\rm fb}^{-1}$, and $\sigma_{hid} > 0.2$ fb for $1000\,{\rm fb}^{-1}$.  We interpret these results in terms of the introduced effective charges.  

Our paper is organized as follows.  In Section~\ref{sigback} we explain our choice of invisible decay channel, and discuss backgrounds.  In Section~\ref{study}, we subject signal and background to cuts to isolate invisible decays, and parametrize the cross section in terms of $Z$ initial state radiation (ISR) and final state radiation (FSR) contributions; hidden decays, which do not couple to the Standard Model, only appear in ISR contributions.  We examine typical masses and couplings that can be probed, as well as kinematic differences between ISR and FSR.  In Section~\ref{findhs} we determine whether decays to hidden sector states can be determined apart from SM neutrinos, using predictions from on-peak data as a background.  Finally, we conclude in Section~\ref{conc}.

\section{Signal and backgrounds} \label{sigback}

We begin by explaining how we search for invisible $Z'$ decays.  We focus on the channel $pp \to ZZ' \to \ell^+\ell^- \met$, where $\ell=e,\mu$.  Other possible 
signal processes to consider are $pp \to \gamma Z' \to \gamma \met$ and $pp \to jZ' \to j \met$.  These, however, are more sensitive to 
uncertainties such as jet energy mismeasurements and jets faking photons.  They require a detailed simulation beyond the scope of our analysis.  We impose the 
following basic acceptance cuts on 
the two leptons in our signal: $|\eta_\ell| < 2.5$, $\Delta R_{\ell\ell} > 0.4$, and $p_T^{\ell} > 10 \, {\rm GeV}$.  We compute the signal using 
MadEvent~\cite{Maltoni:2002qb}; unless noted otherwise, we use MadEvent for all signal and background calculations.

The dominant Standard Model backgrounds to our signal fall into two categories: the production of leptons and neutrinos without a $Z'$ in the intermediate 
state, and the production of $Z+{\rm jets}$ where the jets escape down the beam-pipe or have their energies mismeasured.  We first consider SM production 
of leptons and neutrinos, $pp \to \ell^+\ell^- \nu\bar{\nu}$.  We compute the full SM background with all interference effects and spin correlations 
included.  The primary subprocesses contributing to this background are $pp \to W^+W^- \to 
\ell^+\ell^- \nu\bar{\nu}$ and $pp \to ZZ \to \ell^+\ell^- \nu\bar{\nu}$.  We reduce the $WW$ background using an invariant mass cut on the two leptons: 
$m_Z-10\mbox{ GeV} < m_{\ell \ell} < m_Z + 10 \mbox{ GeV}$.  This restriction helps, but the $WW$ background is still significant.  Further reduction of 
this and the $ZZ$ background is obtained by a $\met$ cut, which we discuss in detail later.  Other kinematic properties, such as the $\Delta\phi$ 
separation between the two leptons in our signal, do not significantly help once the $\met$ cut is imposed.

We must also discuss the potentially large background $pp \rightarrow Z + {\rm jets}$, where the jets escape detection and fake a source of missing $E_T$.  
The LHC hadron calorimeters have a very wide rapidity coverage, up to $\eta \sim 4.9$, but soft jets in the central region are difficult to measure.  We 
therefore restrict ourselves to vetoing jets with $p_T > 50$ GeV in the central region.  Many soft jets may add up to substantial missing $E_T$; 
this can be a problem since the $Z$ cross section is so large to begin with.

We perform a crude estimate of the two possible sources of $Z+{\rm jets}$ background: jets escaping down the beam-pipe or soft jets in the central region.  
We anticipate in this analysis the missing $E_T$ cuts we will later impose to study 1-2 TeV $Z'$ bosons, $\met > 150-200\, {\rm GeV}$.  We begin by 
estimating the cross section for a hard jet with $p_T > 50$ GeV to escape down the beam-pipe.  Using MadEvent we find $81$ ab for $\met > 100$ GeV 
and $29$ ab for $\met > 200$ GeV.  These are very small compared to other backgrounds and will be neglected later in our study.  For softer jets, 
in order to achieve enough missing $E_T$, it will take more (and potentially softer) jets than MadEvent can handle.  We roughly estimate this 
background in the following way.  We require a $Z$ boson and at least one hard jet with $p_T > 30$ GeV in MadEvent.  This cross section is 
$\sim 334$ pb.  The resulting events are then showered using Pythia~\cite{Sjostrand:2006za}.  The surviving cross section drops off rapidly with a 
missing $E_T$ cut:$\sim 690$ ab remains after a cut of $\met > 150$ GeV, and $\lesssim 50$ ab (no generated events remain) after $\met > 200$ GeV.  
We require $\met > 150$ GeV in our analysis, even when a smaller cut would be optimal for the neutrino background, to avoid the $Z+{\rm jets}$ background.  
At this point the background is dominated by the neutrino component, and $Z+{\rm jets}$ can be dropped for statistical purposes.  However, an analysis 
should be performed once $Z + {\rm jets}$ can be determined more precisely.

All signal and background processes in our study are calculated at leading order in the QCD perturbative expansion using a running scale for 
$\alpha_s$.  The next-to-leading order corrections to the background processes are known~\cite{NLOrefs}, while the corrections to the signal process 
are easily calculable.  Since we later use as our significance estimator the ratio of signal over background fluctuation, $S/\sqrt{B}$, we feel this is 
a conservative approach; including the $K$-factors for both $S$ and $B$ would improve our results.  We note that the dependence of the 
next-to-leading order cross section on the renormalization and factorization scales indicates that uncertainties arising from uncalculated higher 
order corrections are at the few percent level or less.  In most of our analysis we also neglect errors 
associated with imprecise knowledge of parton distribution functions.  For the gauge boson production processes considered here, it is likely that LHC data can determine these to high accuracy.  The analysis in Ref.~\cite{Dittmar:1997md} indicates that the parton distribution function errors for di-boson process 
such as the $pp \to WW,ZZ$ backgrounds considered here may be reduced to the percent level by normalizing their rates to the LHC Drell-Yan 
data samples.  Detector effects such as smearing were determined to have a small effect on lepton distributions in \cite{logan}, and we neglect them in our analysis as well.  We neglect other detector issues such as 
smearing of the $\met$ distributions caused by un-vetoed soft jets and the underlying event; although we expect them to be relatively unimportant due to our large missing $E_T$ cut, they are difficult to estimate with current tools.  These issues should be revisited 
in a more complete study which makes use of LHC data, but we believe their neglect is justified in this initial analysis.

\section{Studying the invisible $Z'$} \label{study}

\begin{figure}[htbp]
   \centering
   \includegraphics[width=0.50\textwidth,angle=90]{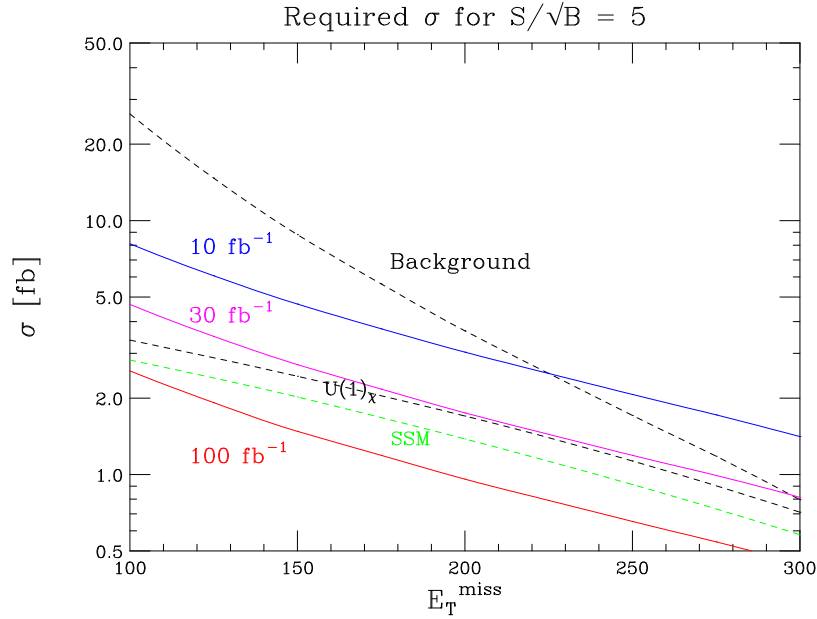}
   \caption{Missing $E_T$ dependence of the SM background and two example $Z'$ models.  Included are curves showing the required $Z'$ cross section 
   for $S/\sqrt{B}=5$ at the LHC for $10\,{\rm fb}^{-1}$, $30\,{\rm fb}^{-1}$, and $100\,{\rm fb}^{-1}$.}
   \label{5s}
\end{figure}

\begin{figure}[htbp]
   \centering
   \includegraphics[width=0.50\textwidth,angle=90]{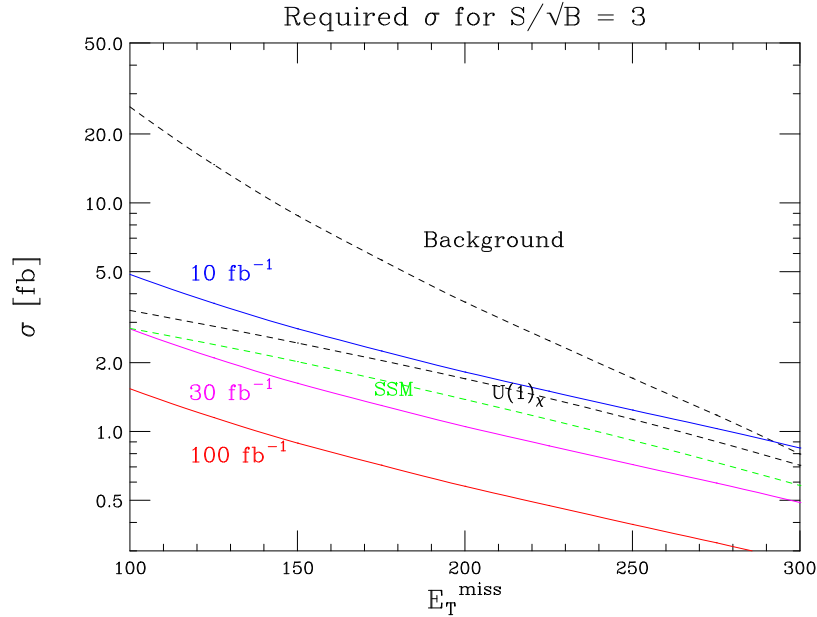}
   \caption{Missing $E_T$ dependence of the SM background and two example $Z'$ models.  Included are curves showing the required $Z'$ cross section 
   for $S/\sqrt{B}=3$ at the LHC for $10\,{\rm fb}^{-1}$, $30\,{\rm fb}^{-1}$, and $100\,{\rm fb}^{-1}$.}
   \label{3s}
\end{figure}

Employing the cuts and techniques described in the previous section, we determine whether invisible $Z'$ decays can be observed over the SM background.  
We map out the missing $E_T$ dependence in Figs.~(\ref{5s}) and~(\ref{3s}).  The basic cuts outlined in the previous section have been implemented.  
Both plots show the SM background $pp \to \ell^+\ell^- \nu\bar{\nu}$ as a function of a lower cut on the missing $E_T$.  Two fiducial models are also shown: the sequential Standard Model and the 
$U(1)_\chi$ model with an overall gauge coupling $g' = 1$.  We assume $M_{Z'}=1$ TeV for both models.  The plots begin at $\met = 100$ GeV to avoid 
serious issues with the $Z + {\rm jets}$ background.  We also plot the required invisible $Z'$ cross sections for observation at the LHC assuming 
$10\,{\rm fb}^{-1}$, $30\,{\rm fb}^{-1}$, and $100\,{\rm fb}^{-1}$.  Fig.~(\ref{5s}) shows the required cross section for a statistical significance of 
$S/\sqrt{B}=5$, while Fig.~(\ref{3s}) shows the required rate for $S/\sqrt{B}=3$.  Two facts can be observed from these graphs.  First, the optimum 
missing $E_T$ cut for TeV mass $Z'$ bosons is around 200 GeV, above the level where $Z + {\rm jets}$ is a serious concern.  Second, for realistic models a 
signal is observable at the LHC even with moderate integrated luminosity.  $S/\sqrt{B}=3$ is possible for both fiducial models with less than 
$30\,{\rm fb}^{-1}$, while $S/\sqrt{B}=5$ is possible for less than $100\,{\rm fb}^{-1}$.


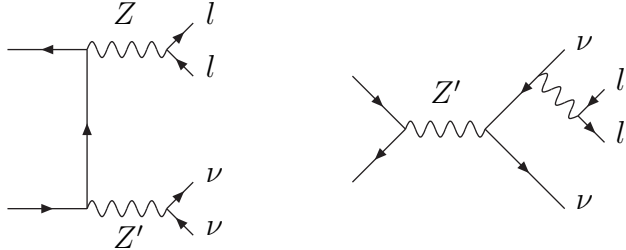
\begin{figure}[htb]
  \begin{center}
    \begin{picture}(250,100)(0,0)
      \SetColor{Black}
        \ArrowLine(0,10)(30,10)
        \ArrowLine(30,10)(30,70)
        \ArrowLine(30,70)(0,70)
        \Photon(30,70)(60,70){3}{4}
        \Photon(30,10)(60,10){3}{4}
        \put(40,80){$Z$}
        \put(40,-5){$Z'$}
        \ArrowLine(60,10)(70,20)
        \ArrowLine(70,0)(60,10)
        \ArrowLine(60,70)(70,80)
        \ArrowLine(70,60)(60,70)
        \put(75,80){$l$}
        \put(75,60){$l$}
        \put(75,0){$\nu$}
        \put(75,20){$\nu$}

        \ArrowLine(130,60)(150,40)
        \ArrowLine(150,40)(130,20)
        \Photon(150,40)(180,40){3}{4}
        \ArrowLine(210,70)(180,40)
        \ArrowLine(180,40)(210,10)
        \Photon(200,60)(215,45){3}{3}
        \ArrowLine(215,45)(225,35)
        \ArrowLine(225,55)(215,45)
        \put(160,50){$Z'$}
        \put(230,35){$l$}
        \put(230,55){$l$}
        \put(215,70){$\nu$}
        \put(215,10){$\nu$}

    \end{picture}
  \end{center}
\caption{\label{IFdiags} 
          Example initial-state radiation diagram (left) and final-state radiation diagram (right).  The particle labeled $\nu$ can denote either a SM 
          neutrino or hidden sector state; in the second case, it can only be produced via initial-state radiation.}
\end{figure}

We wish to do more than simply observe the invisibly decaying $Z'$.  We also want to measure the underlying parameters leading to these decays, and 
determine whether the decays are accounted for by SM neutrinos only, or whether decays to other exotic states are occuring.  Although at first this appears more model-dependent, the matrix element for $ZZ'$ 
production actually possesses a simple structure that can be encapsulated in a few quantities.  Two distinct classes of Feynman diagrams contribute to the 
process: final-state radiation (FSR) graphs where the $Z$ is 
emitted from the neutrinos, and initial-state radiation (ISR) graphs where the $Z$ is emitted from the initial quark line.  Examples of each type are shown 
in Fig.~(\ref{IFdiags}).  We note that because of the invariant mass cut, diagrams where the leptons are emitted from the $Z'$ are 
numerically negligible.  The particle labeled $\nu$ in the graphs can denote either a SM neutrino or a hidden sector state.  If it is a hidden state, it does not 
couple to the $Z$ boson and therefore can be produced only via ISR graphs.  We have checked that the interference of ISR and FSR contributions is numerically 
small, indicating that only squared ISR and squared FSR graphs contribute to the signal cross section.  This can be partially understood by noting that the $Z'$ 
propagator cannot be simultaneously on-shell in both types of diagrams, indicating that for narrow states the interference should be suppressed.

\begin{figure}[htbp]
   \centering
   \includegraphics[width=0.5\textwidth,angle=90]{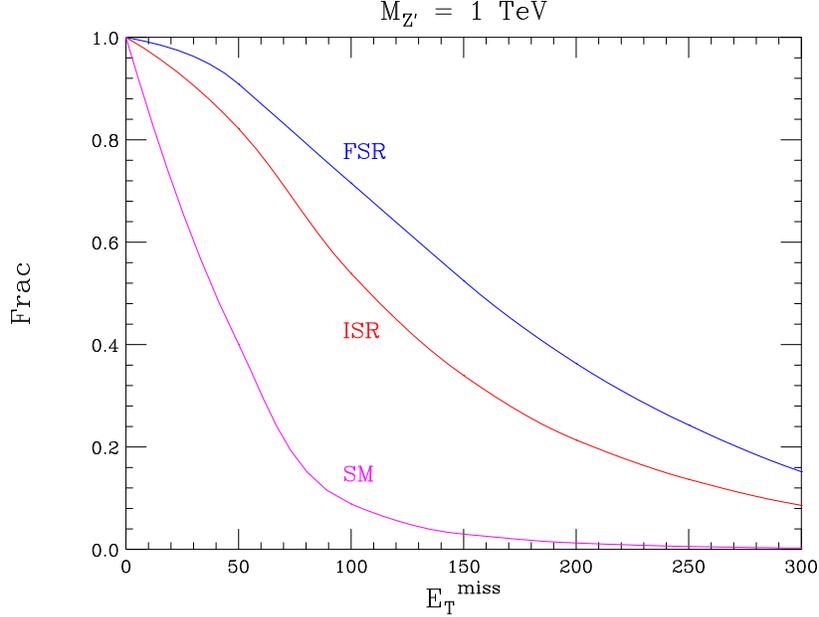}
   \caption{\label{ISRFSR} Fractions of ISR, FSR, and SM events which survive a lower missing $E_T$ cut.}
\end{figure}

Generically, an ISR $Z$ will be softer than one from FSR, so that we can expect a corresponding preference for a softer $\met$ spectrum from ISR than FSR.  
This is shown in Fig.~(\ref{ISRFSR}), where the fraction of the total ISR or FSR cross-section surviving a given $\met$ cut is shown.  It is seen that the 
ISR contribution drops off more quickly, as expected.  Also shown is the SM background, which drops off more quickly than either $Z'$ contribution.

The relative size of the ISR and FSR contributions determines how well a $Z'$ decaying to hidden sector particles can be extracted.  A large ISR contribution 
implies that non-standard decays can be measured.  The simplicity of the matrix element structure allows us to parametrize how different $Z'$ states 
decay via ISR and FSR contributions in a model independent way.  To see this, we first write the cross section subject to the basic acceptance cuts 
and missing $E_T$ cut as
\be
\sigma = \sigma_{ISR}^u + \sigma_{ISR}^d + \sigma_{FSR}^u + \sigma_{FSR}^d,
\ee  
where up and down quark contributions have been separated.  Each $\sigma_{ISR,FSR}^{u,d}$ can in turn be written as a product of two distinct terms: 
a piece which 
incorporates the matrix elements, parton distribution functions, and experimental cuts, denoted as $f_{ISR,FSR}^{u,d}$; a piece which depends on the charges 
from a given model, $Q_{ISR,FSR}^{u,d}$.  We then have $\sigma_{ISR,FSR}^{u,d} = f_{ISR,FSR}^{u,d}Q_{ISR,FSR}^{u,d}$.  The coupling structure of the 
various terms takes the form
\be
Q_{ISR}^{q} \equiv \left((q_V'^2+q_A'^2)(q_V^2+q_A^2)+4 q_V' q_A' q_V q_A \right) \frac{\Gamma_{Z'}^{inv}}{\Gamma_{Z'}}
\ee
and
\be
Q_{FSR}^q \equiv (q_V'^2+q_A'^2)(q_V^2+q_A^2)\frac{\Gamma_{Z'}^{SM\nu}}{\Gamma_{Z'}},
\ee
where $\Gamma_{Z'}^{SM\nu}$, $\Gamma_{Z'}^{inv}$ denote the partial widths of the $Z'$ to SM $\nu$'s or to {\em any} invisible particle (SM $\nu$'s 
or hidden sector states), and $\Gamma_{Z'}$ is the total width.  A prime on a charge indicates that it is a $Z'$ charge, while no prime 
denotes a SM $Z$ charge.  $A$ and $V$ subscripts denote axial and vector charges, respectively.  
Any $Z'$ model can then be constructed by dialing $Q_{ISR,FSR}^{u,d}$ appropriately.  The functions $f_{ISR,FSR}^{u,d}$ depend on the given model 
under consideration only through the $Z'$ mass in the narrow width approximation.  

Values of the $Q$ charges are given in Table~(\ref{Qcharge}) for the the sequential Standard Model (SSM) and $U(1)_{\chi}$ model discussed previously.  
We also show a $U(1)_B$ model with gauge coupling $g'=1$ in which the $Z'$ couples to baryon number, and which also includes a hidden sector state, 
assumed to be a vector-like fermion with unit charge.  We present charge values for the SSM and $U(1)_{\chi}$ models with the same hidden state.  
The increase of the $Q_{ISR}$ when the hidden state is present can be observed in Table~(\ref{Qcharge}).  We will see later that the 
$Q$ values assuming only SM neutrino decays are determined once the Drell-Yan channel $pp \to Z' \to \ell^+\ell^-$ is observed.  Measuring different 
$Q$ values than predicted by Drell-Yan studies would indicate the presence of hidden sector $Z'$ decays.  If leptonic $Z'$ decays do not 
occur, such as in the $U(1)_B$ model, the $pp \to ZZ' \to \ell^+\ell^- \met$ process considered here becomes a discovery channel.

\begin{table}
\centering
\begin{tabular}{|c|c|c|c|c|c|} \hline
& $U(1)_{\chi}$ & $U(1)_{\chi}^{hid}$  & SSM & SSM$^{hid}$ & $U(1)_B$ \\ \hline
$Q_u^{FSR}$ &0.274 &0.212 &0.292 &0.197 &0 \\
$Q_d^{FSR}$ &1.75 &1.36 &0.481 &0.324 &0 \\
$Q_u^{ISR}$ &0.274 &0.589 &0.436 &1.08 &1.49 \\
$Q_d^{ISR}$ &0.432 &0.931 &0.907 &2.26 &1.90 \\ \hline
$u_V$ & 0 & 0 & $\frac{1}{4}-\frac{2}{3}\sin^2\theta_W$ & $\frac{1}{4}-\frac{2}{3}\sin^2\theta_W$ & $\frac{1}{3}$\\
$u_A$ & $\frac{1}{2\sqrt{6}}$ & $\frac{1}{2\sqrt{6}}$ & $\frac{-1}{4}$ & $\frac{-1}{4}$ & 0 \\
$d_V$ & $\frac{-2}{\sqrt{6}}$ & $\frac{-2}{\sqrt{6}}$ & $\frac{-1}{4}+\frac{1}{3}\sin^2\theta_W$ & $\frac{-1}{4}+\frac{1}{3}\sin^2\theta_W$ & $\frac{1}{3}$ \\
$d_A$ & $\frac{-1}{\sqrt{6}}$ & $\frac{-1}{\sqrt{6}}$ & $\frac{1}{4}$ & $\frac{1}{4}$ & 0 \\
$e_V$ & $\frac{2}{\sqrt{6}}$ & $\frac{2}{\sqrt{6}}$ & $\frac{-1}{4}+\sin^2\theta_W$ & $\frac{-1}{4}+\sin^2\theta_W$ & 0 \\
$e_A$ & $\frac{-1}{\sqrt{6}}$ & $\frac{-1}{\sqrt{6}}$ & $\frac{1}{4}$ & $\frac{1}{4}$ & 0 \\
$X_V$ & 0 & 1 & 0 & 1 & 1 \\
$X_A$ & 0 & 0 & 0 & 0 & 0 \\ \hline

\end{tabular}
\caption{\label{Qcharge}$Q$'s for various models, multiplied by $10^3$.  We have also included the underlying charges of the considered model for orientation, with hidden state charges denoted by $X$.  See the text for further explanation.}
\end{table}

\begin{figure}[htbp]
   \centering
   \includegraphics[width=0.5\textwidth,angle=90]{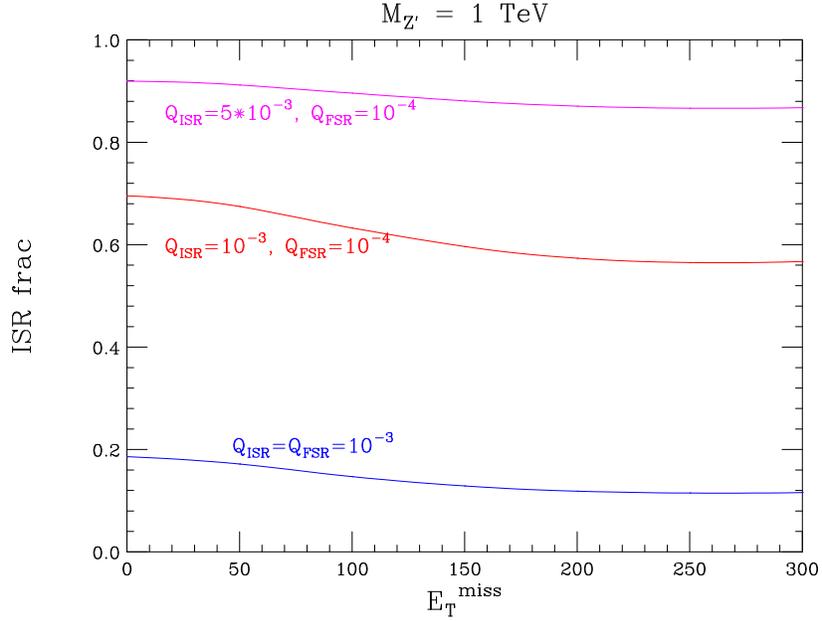}
   \caption{Fraction of cross section coming from ISR initiated diagrams as a function of missing $E_T$ cut for three example $Q$ choices.}
   \label{ISRfrac}
\end{figure}

\begin{figure}[htbp]
   \centering
   \includegraphics[width=0.5\textwidth,angle=90]{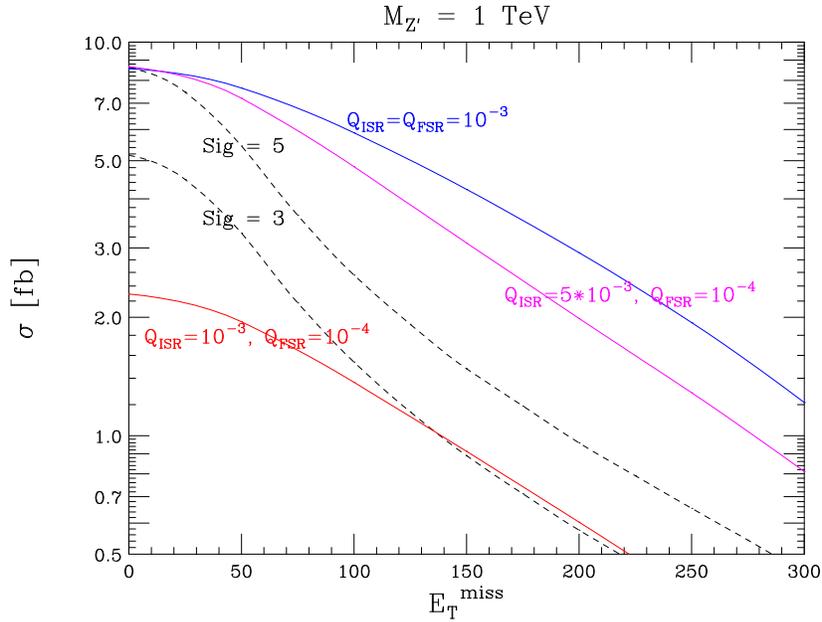}
   \caption{Cross section as a function of missing $E_T$ cut for three example $Q$ choices.  The cross sections required for $S/\sqrt{B}=3,5$ 
     assuming $100 \mbox{ fb}^{-1}$ are shown as dashed lines.}
   \label{absXsct}
\end{figure}

To develop some intuition, we present below several plots showing features of the cross section for different $Q$ choices.  For simplicity of 
presentation we make the simplifying assumption $Q_{FSR}^u=Q_{FSR}^d=Q_{FSR}$ and $Q_{ISR}^u=Q_{ISR}^d=Q_{ISR}$.  The degeneracy between $Q^u_{ISR,FSR}$ and $Q^d_{ISR,FSR}$ in the plots can be broken by utlizing the information $f_{ISR}^u = 353$ fb, $f_{ISR}^d = 227$ fb, $f_{FSR}^u = 2.71$ pb, $f_{FSR}^d = 1.40$ pb, evaluated for a missing $E_T$ cut of 150 GeV.  We note that the kinematic 
dependences of the $u$ and $d$-quark cross sections on the missing $E_T$ cut are very similar.  We focus on three example cases: 
$Q_{FSR}=Q_{ISR}=10^{-3}$; $Q_{FSR}=10^{-4}$ and $Q_{ISR}=10^{-3}$; $Q_{FSR}=10^{-4}$ and $Q_{ISR}=5\times 10^{-3}$.  These values are roughly 
consistent with those present in typical models as shown in Table~(\ref{Qcharge}).  
We show in Fig.~(\ref{ISRfrac}) the ISR fraction of the total cross section as a function of the missing $E_T$ cut for $M_{Z'}=1$ TeV.  For 
$Q_{FSR}=Q_{ISR}$, the ISR fraction of the cross section is less than 20\%.  The FSR matrix elements give a larger contribution to the cross section, 
suggesting that it will be difficult to dig out the hidden sector sector component from invisible decays.  We will quantify this further later.  The cross 
sections for the $Q$ charges under consideration are shown in Fig.~(\ref{absXsct}).  For comparison, we overlay the curves showing the required 
cross sections for $S/\sqrt{B}=3,5$ with $100 \mbox{ fb}^{-1}$ from Figs.~(\ref{5s}) and ~(\ref{3s}).  We see that at least $S/\sqrt{B}=3$ evidence 
is possible at the LHC for a range of $Q$ values.

\begin{figure}[htbp]
   \centering
   \includegraphics[width=0.5\textwidth,angle=90]{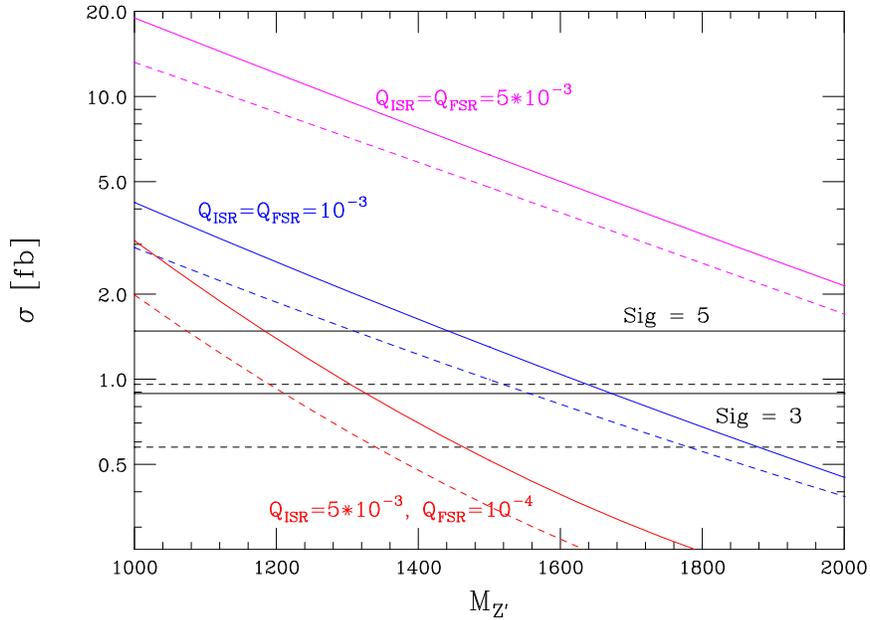}
   \caption{The $Z'$ cross section as a function of mass for several different example $Q$ values.  The solid lines assume a cut $\met>150$ GeV, the dashed 
     lines assume $\met>200$ GeV.  The horizontal lines are the cross sections required for $S/\sqrt{B}=3,5$ with $100 \mbox{ fb}^{-1}$; again, solid 
     lines assume $\met>150$ GeV and dashed lines assume $\met>200$ GeV.  The mass reach for a given missing $E_T$ cut and signifance is determined 
     by finding the appropriate intersection of curve and horizontal line.}
   \label{Massdep}
\end{figure}

To study what $Z'$ masses can be probed, we show in Fig.~(\ref{Massdep}) the $Z'$ cross section as a function of mass for several different example $Q$ values.  
Since the masses are larger, the corresponding $Q$ values needed for observation are larger, so we present results assuming somewhat larger charges.  We show the 
results for missing $E_T$ cuts of both 150 and 200 GeV; the value that actually maximizes $S/\sqrt{B}$ varies with $M_{Z'}$.  Included in this plot are the 
required cross sections for $S/\sqrt{B}=3,5$ assuming $100 \mbox{ fb}^{-1}$.  For $Q_{FSR}=5 \times 10^{-3}$, masses beyond 2 TeV are easily observable.  If 
$Q_{FSR}=10^{-4}$ and the ISR charge is larger, the case relevant for $Z'$ decays to hidden sectors, only masses up to 1.25 or 1.5 TeV can be probed with 
$S/\sqrt{B}=5$.

Finally, if the $Z'$ does not decay into leptons but does decay to hidden sector states, $pp \to ZZ' \to \ell^+\ell^- \met$ becomes a discovery channel.  The 
experimental search for this leptophobic $Z'$ will proceed by moving upward a minimum missing $E_T$ cut and looking for a signal to emerge.  The shape 
of the $\met$ spectrum should give some sensitivity to the $Z'$ mass.  Also, if a more complicated structure of new physics than a simple isolated 
$Z'$ is discovered, we will want to determine whether the $\ell^+\ell^- \met$ signal arises from a single new gauge boson or multiple states.  

\begin{figure}[htbp]
   \centering
   \includegraphics[width=0.5\textwidth,angle=90]{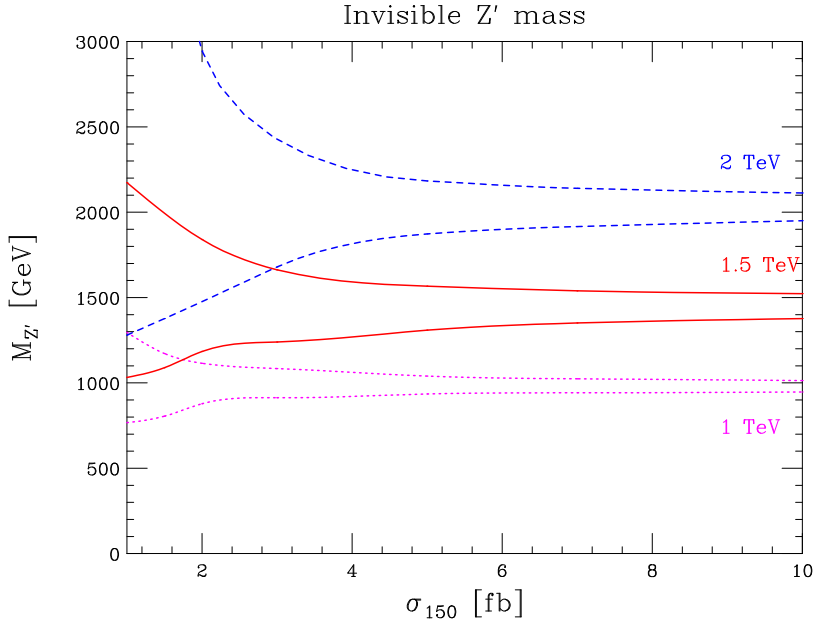}
   \caption{$1 \sigma$ statistical error bands on $Z'$ mass measurement, given hypotheses of $M_{Z'}=1,1.5,2$ TeV, as a function of the total cross-section $pp \rightarrow ZZ' \rightarrow \ell^+ \ell^- \met$ with $\met > 150 \mbox{ GeV}$.}
   \label{merrs}
\end{figure}

We determine the statistical measurement error for three fiducial $Z'$ masses, $1, 1.5$, and $2$ TeV, by performing a $\chi^2$ comparison of their missing $E_T$ spectra versus other masses.  We set $Q_{FSR}=0$ to simulate a completely leptophobic $Z'$ for our spectra.  The cross section is divided into several bins in missing $E_T$; we take 
the ratio of each bin to the total rate surviving the $\met > 150$ cut to normalize.  We generate $\met$ templates for many other masses and compare the ratios in each bin 
to the ratios for each fiducial mass, and determine for what masses a total 1 $\sigma$ deviation is exceeded in each case; this occurs when the total $\chi^2$ reaches $1$.  In Fig.~(\ref{merrs}) we have plotted 1 $\sigma$ error bands for the three $Z'$ masses as a function of hidden cross section after the missing $E_T$ cut of $150$ GeV, for the SLHC luminosity of $1 \mbox{ ab}^{-1}$.  We have not taken into account errors other than statistical, such as PDF uncertainties; we leave the inclusion of such errors for a more complete analysis.  One can see, however, that given just the statistical error, there is good sensitivity to the mass given a sufficiently large cross section (reasonable for a leptophobic $Z'$) and sufficient integrated luminosity.  For reference, a $1$ fb, $\met > 150$ cross section corresponds to $Q_{ISR}^u = Q_{ISR}^d$ values of $0.00172$, $0.0123$, and $0.0377$, for masses of $1$ TeV, $1.5$ TeV, and $2$ TeV, respectively. 

\section{Finding the hidden sector} \label{findhs}

We wish to study whether LHC results can determine if invisible $Z'$ decays occur only to SM neutrinos, or whether other states are involved.  This would provide 
insight into possible hidden sectors to which the $Z'$ couples.  

The crucial fact that allows this measurement to be performed is that the charges $Q$ introduced in the previous section are predicted by the analysis of Drell-Yan $Z'$ production 
in~\cite{Petriello:2008zr} if the $Z'$ decays invisibly only to neutrinos.  We note that 
\be
Q_{ISR}^q = \left( \frac{c_q}{2} \frac{C}{C+1} (q_V^2+q_A^2) + e_q \frac{C}{C-1} q_V q_A \right) \frac{\Gamma_{Z'}^{inv}}{\Gamma_{Z'}^{\nu}},
\label{QISRonpeak}
\ee
and
\be
Q_{FSR}^q = \frac{c_q}{2} \frac{C}{C+1} (q_V^2+q_A^2), 
\label{QFSRonpeak}
\ee  
where $c_q$ and $e_q$ are the on-peak couplings determined in~\cite{Petriello:2008zr}:
\begin{eqnarray}
c_q&=&\frac{M_{Z'}}{24\pi\Gamma}(q_R^{\prime 2}+q_L^{\prime 2})(l_R^{\prime 2}+l_L^{\prime 2}); \nonumber \\ 
e_q&=&\frac{M_{Z'}}{24\pi\Gamma}(q_R^{\prime 2}-q_L^{\prime 2})(l_R^{\prime 2}-l_L^{\prime 2}); \nonumber \\ 
C&=&\frac{l_L^{\prime 2}}{l_R^{\prime 2}} = \frac{c_u+e_u-c_d-e_d}{c_u-e_u-c_d+e_d}.
\end{eqnarray}
$Q^q_{FSR}$ is fixed by $c_q$ and $e_q$.  If the $Z'$ decays invisibly only to neutrinos, then $\Gamma_{Z'}^{inv} = \Gamma_{Z'}^{\nu}$; $Q^q_{ISR}$ is then completely predicted by the on-peak couplings.  Any deviation of $Q^q_{ISR}$ from this limit indicates additional invisible decays of the $Z'$.

\begin{figure}[htbp]
   \centering
   \includegraphics[width=0.5\textwidth,angle=90]{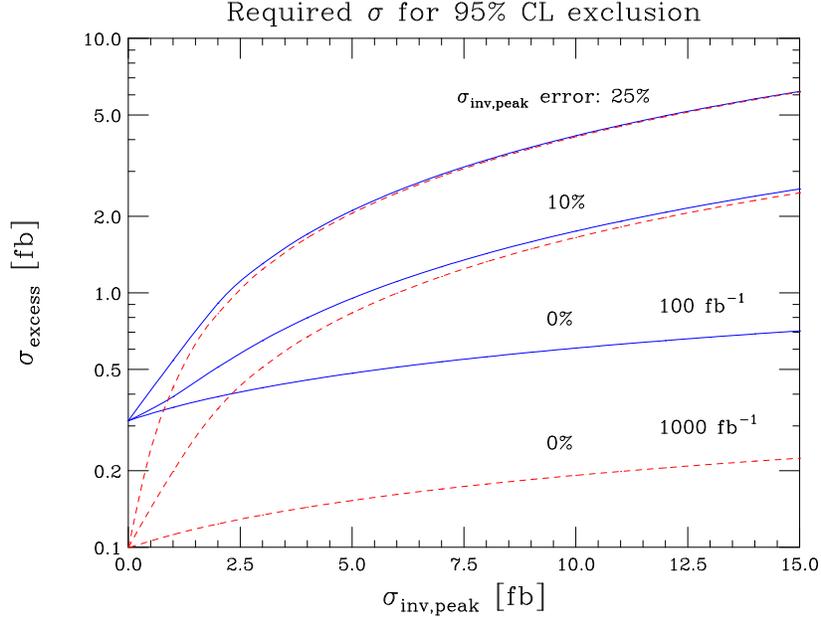}
   \caption{Excess cross section over that predicted by on-peak data, $\sigma_{inv}^{peak}$, which can be excluded at 95\% confidence level  
for $100 \mbox{ fb}^{-1}$ and 
$1000 \mbox{ fb}^{-1}$.  Errors on the predicted $\sigma_{inv}^{peak}$ of 0\%, 10\% and 25\% from on-peak data are assumed.  $\sigma_{excess}$ results from decays to 
hidden sector particles.}
   \label{excess95}
\end{figure}

\begin{figure}[htbp]
   \centering
   \includegraphics[width=0.5\textwidth,angle=90]{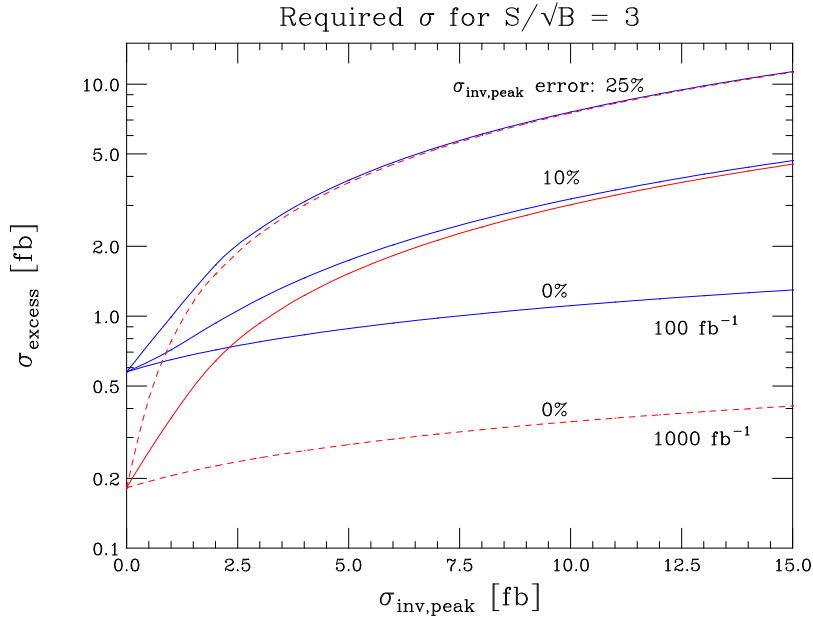}
   \caption{Excess cross section over that predicted by on-peak data, $\sigma_{inv}^{peak}$, which can be observed at $3\sigma$ for $100 \mbox{ fb}^{-1}$ and 
     $1000 \mbox{ fb}^{-1}$.  Errors on the predicted $\sigma_{inv}^{peak}$ of 0\%, 10\% and 25\% from on-peak data are assumed.}
   \label{excess3}
\end{figure}

We first determine how big an excess over the expected invisible cross section predicted by on-peak data can be observed.  In addition to SM production of leptons and missing energy, the signal $pp \rightarrow Z Z' \rightarrow \ell^+ \ell^- \nu_\ell \bar{\nu}_\ell$ now becomes a background to $pp \rightarrow Z Z' \rightarrow \ell^+ \ell^- X \bar{X}$, where the $X$s are the hidden sector particles.  In Fig.~(\ref{excess95}) we show the the size of the excess cross section over that predicted by the on-peak data which can be excluded at 95\% C.L. for $100 \mbox{ fb}^{-1}$ and $1000 \mbox{ fb}^{-1}$ of data; as this is a difficult measurement we have assumed a sizeable amount of integrated luminosity.  The 
excess cross section for which $3 \sigma$ evidence can be obtained is shown in Fig.~(\ref{excess3}).  We have used a cut of $\met>200$ GeV in producing these numbers.  The excess 
cross section that can be probed depends crucially on how well the expected invisible cross section can be predicted from on-peak data.  To determine this precision, 
the expected errors on $c_q,e_q$ from~\cite{Petriello:2008zr} must be propagated through the expressions in Eqs.~(\ref{QISRonpeak}) and~(\ref{QFSRonpeak}).  We present 
results for fractional errors on the predicted invisible cross section of 10\% and 25\%, which are consistent with the error propagation, as well as for the idealized limit of no error.  From Figs.~(\ref{absXsct}) and~(\ref{Massdep}), we see that cross sections for typical $Q$ values with $\met>200$ GeV are between 1-10 fb.  Using the 
10\% error curve from Fig.~(\ref{excess95}), hidden sector decays leading to excess cross sections of 1-2 fb can be excluded at 95\% confidence level.  If no on-peak 
cross section is observed, then the left side of Figs.~(\ref{excess95}) and~(\ref{excess3}) indicate how well completely invisibly decaying $Z'$ bosons can be probed.  
Completely invisibly decaying $Z'$ boson cross sections can be excluded down to 0.5 fb given sufficient integrated luminosity.  

Several reductions of the error associated with the invisible cross section prediction are possible.  With $100 \, {\rm fb}^{-1}$ the error comes mostly from 
parton distribution functions; with $1000 \, {\rm fb}^{-1}$ it comes entirely from parton distribution functions.  These uncertainties will be 
significantly improved with LHC data.  In addition, one may be able to normalize the FSR contribution to on-peak data, due to the similar PDF and coupling structure.  Approaching a 5\% error is not inconceivable.

\begin{figure}[htbp]
   \centering
   \includegraphics[width=0.5\textwidth,angle=90]{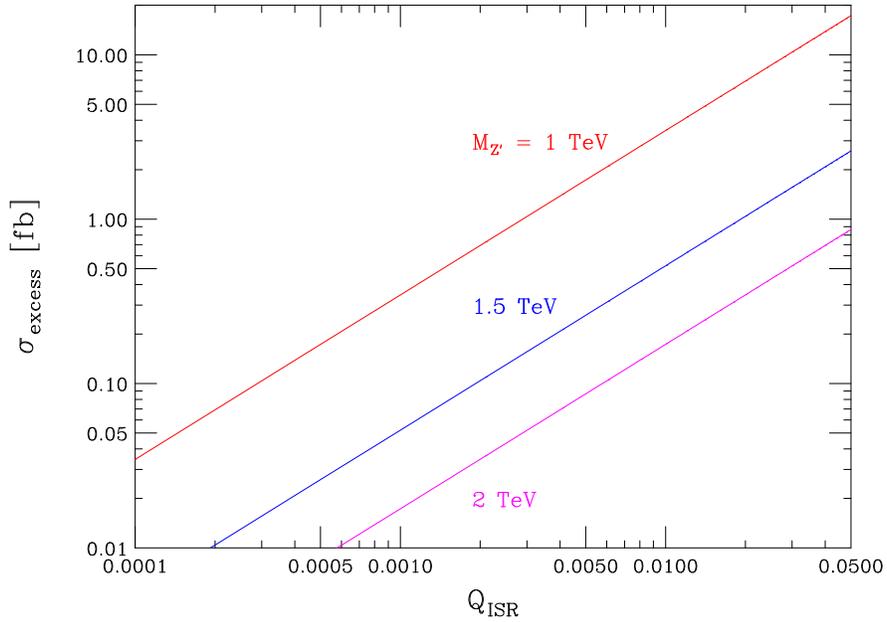}
   \caption{Excess cross section $\sigma_{excess}$ as a function of $Q_{ISR}^{u,d} = Q_{ISR}$.}
   \label{excessISR}
\end{figure}

We now interpret this excess cross section using our effective charges.  We write $Q_{ISR}^{u,d} = Q_{SM \nu}^{u,d}+Q_{hid}^{u,d}$, where $Q_{SM \nu}$ can be 
predicted from the on-peak data and $Q_{hid}^{u,d}$ is the portion coming from decays to hidden sector states.  We plot in Fig.~(\ref{excessISR}) the size 
of this excess cross section as a function of $Q_{hid}$, where $Q_{hid}^u = Q_{hid}^d$.  Using this graph and keeping in mind the 1-2 fb cross sections, we observe 
that it will be difficult to significantly constrain hidden sector decays if $M_{Z'}$ is significantly greater than 1 TeV.  For a 1 TeV state, charges in the range 
$Q_{hid}^q \geq 5 \times 10^{-3}$ can be probed.

Although the 95\% confidence level and $3\sigma$ reaches in the $Q-M_{Z'}$ plane can be determined from Figs.~(\ref{absXsct}),~(\ref{Massdep}),~(\ref{excess95}),~(\ref{excess3}), and~(\ref{excessISR}), since 
the parameter space is large and the graphs are numerous, we summarize below several canonical cases.

\begin{itemize}

\item \underline{95\% exclusion for pure hidden sector $Z'$}: From Fig.~(\ref{excess95}), the required cross sections to exclude this state are 
  $\sigma_{excess}>0.3$ fb with $100 \,{\rm fb}^{-1}$ and $\sigma_{excess}>0.1$ fb with $1000 \,{\rm fb}^{-1}$.  This implies the following 
  exclusion limits for fixed $M_{Z'}$, $Q_{hid}$. 
  \begin{itemize}
     \item $M_{Z'}=1$ TeV: $Q_{hid} < 2\times 10^{-3}$ with $100 \,{\rm fb}^{-1}$ and 
              $Q_{hid} < 5\times 10^{-4}$ with $1000 \,{\rm fb}^{-1}$ using  Fig.~(\ref{excessISR}).
    \item $Q_{hid}=5 \times 10^{-3}$: $M_{Z'} > 1300$ GeV with $100 \,{\rm fb}^{-1}$
             and $M_{Z'} > 1700$ GeV with $1000 \,{\rm fb}^{-1}$ using Fig.~(\ref{Massdep}).
 \end{itemize}
  
  \item \underline{$Z'$ boson with $M_{Z'}=1\,{\rm TeV}$, $Q_{ISR}=5 \times 10^{-3}$, $Q_{FSR}=10^{-4}$}: We assume a 10\% error in the invisible cross section 
  prediction when interpreting this state.  Using the graphs in a similar fashion as above, the following information about $Q_{hid}$ can be obtained.
  \begin{itemize}
    \item {\it 95\% exclusion}: $Q_{hid}<2\times 10^{-3}$ with $100 \,{\rm fb}^{-1}$ and $Q_{hid}< 10^{-3}$ with $1000 \,{\rm fb}^{-1}$.
    \item {\it $3\sigma$ evidence}: can probe $Q_{hid}=4\times 10^{-3}$ with $100 \,{\rm fb}^{-1}$ and $Q_{hid}=2\times 10^{-3}$ with $1000 \,{\rm fb}^{-1}$.
  \end{itemize}

\end{itemize}

\section{Conclusions} \label{conc}

We have studied the feasibility of observing an invisible $Z'$ through the process $pp \rightarrow Z Z'\rightarrow \ell^+ \ell^- X X^\dagger$ at the LHC, where $X$ is any neutral, (quasi-) stable state.  We found that $3 \sigma$ evidence of this process could be made with as little as $10 \mbox{ fb}^{-1}$ of data for a standard $1 \mbox{ TeV}$ $U(1)_\chi$ $Z'$ with gauge coupling $g'=1$, while a $5\sigma$ discovery is possible with $30 \mbox{ fb}^{-1}$.  With our results, using Figs.~(\ref{absXsct}-\ref{Massdep}) in conjunction with Figs.~(\ref{5s}-\ref{3s}), the discovery reach of LHC for observing any invisibly decaying $Z'$ can be computed.  We parametrized our results in terms of two effective charges that completely describe production of a $Z'$ in conjunction with a $Z$ radiated off the initial state (ISR) or the final state (FSR).  We found that for a 1 TeV $Z'$, any model with $Q_{ISR} > 10^{-3}$, $Q_{FSR} > 10^{-4}$ can be observed at $3\sigma$ with $100 \mbox{ fb}^{-1}$ of data.  This shows that a leptophobic $Z'$ that cannot be observed through the usual 
Drell-Yan channel at the LHC can be discovered if it decays invisibly to hidden sector states.  We showed that some sensitivity to the $Z'$ 
mass can be obtained by studying the missing $E_T$ spectrum.

In addition, we demonstrated that an excess invisible decay of the $Z'$ to hidden sector states over the predicted cross section for $pp \rightarrow ZZ' \rightarrow \ell^+ \ell^- \bar{\nu} \nu$ from on-peak data can be excluded at 95\% confidence level if the size of this cross section is 20-30\% of the total cross section, given a 10\% error on the predicted invisible cross section.  The exotic states may, for example, be dark states from a ``Hidden Valley'' model \cite{Strassler:2006im}.  The $Z'$ may be a communicator to a light hidden sector with MeV mass dark matter states as in the model of \cite{Hooper:2008im}; this is motivated by the INTEGRAL/SPI observation \cite{Jean:2003ci} of a 511 keV line toward the galactic center.  
To get 20-30\% of the invisible cross section of the $Z'$ from hidden decays will require in most cases a hidden sector with multiple states to compete with the SM neutrino invisible decays.  In particular, when $Q_{ISR} \simeq Q_{FSR}$, the branching fraction to new hidden sector states must be approximately the same as the branching fraction to SM neutrinos in order to obtain a $\sim 20\%$ deviation in the invisible cross section $pp \rightarrow Z Z'\rightarrow \ell^+ \ell^- X X^\dagger$ predicted from on-peak data.   This is on account of the hidden sector states entering only through graphs where the $Z$ is radiated off the initial state quark lines; these initial state graphs usually compose a relatively small fraction of the total cross section: $\sim 20\%$ from Fig.~(\ref{ISRfrac}) for the fiducial case of the effective charges for initial state and final state $Z$ radiation being roughly the same.  Despite this potential difficulty of observing decays to hidden sector states, we have shown that it is nonetheless feasible, given the presence of such a hidden sector.  The possibility to observe such dark states through the hidden decays of a new vector gauge boson makes the accurate measurement of the invisible $Z'$ at the LHC an exciting and reachable goal.

\section*{Acknowledgments}

The authors are supported by the DOE grant DE-FG02-95ER40896, Outstanding  Junior Investigator Award, 
by the University of Wisconsin Research Committee
with funds provided by the Wisconsin Alumni Research Foundation, and
by the Alfred P.~Sloan Foundation.

\end{document}